\documentclass[12pt]{article}
\usepackage{amsfonts}
\usepackage{amsmath}
\usepackage{amssymb}
\usepackage{graphicx}
\usepackage{color}
\usepackage[all, knot]{xy}
\usepackage{tikz}
\usepackage{array}

\usepackage{ulem}

\usepackage[utf8]{inputenc}
\usepackage{epstopdf}
\usepackage[footnotesize]{caption}
\usepackage{amsthm}
\usepackage{enumitem}
\usepackage{mathrsfs}

\usepackage[margin=3cm]{geometry}

\def \be {\begin{equation}}
\def \ee {\end{equation}}
\def \bea {\begin{eqnarray}}
\def \eea {\end{eqnarray}}
\def \nn {\nonumber}

\def \rr {\raise.35ex\hbox{\small $\prime$}\kern-.17em{\mbox{\large $\imath$}}}

\def \dels {\partial\kern-.6em /\kern.1em}
\def \As {{A\kern-.5em / \kern.5em}}
\def \Ds {D\kern-.7em / \kern.5em}

\def \ks {k\kern-.5em /}
\def \ls {l\kern-.5em /}

\newcommand{\ci}[1]{}
\newcommand{\ba}{\begin{eqnarray}}
\newcommand{\ea}{\end{eqnarray}}
\newcommand{\bal}{\begin{align}}
\newcommand{\eal}{\end{align}}
\newcommand{\bay}[1]{\left(\begin{array}{#1}}
\newcommand{\eay}{\end{array}\right)}

\newcommand{\hide}[1]{}

\newlist{axioms}{enumerate}{2}
\setlist[axioms,1]{label=\textbf{A\arabic{axiomsi}.}, ref=A\arabic{axiomsi}}
\setlist[axioms,2]{label=\textbf{A\arabic{axiomsi}\rlap{\myEnumCounter{axiomsii}}.},%
                   ref=A\arabic{axiomsi}\myEnumCounter{axiomsii},%
                   align=parleft,%
                   leftmargin=0em,%
                   itemsep=1.4ex,%
                   before={\stepcounter{axiomsi}}}

  \usetikzlibrary{decorations.markings}

\begin{document}
\begin{titlepage}
%\begin{flushright}
%NORDITA 2019-102 %\\
%October,  2019
%\end{flushright}

\begin{center}

\textbf{\LARGE
$(p-1)$-Bracket for D$p$-branes in\\ 
Large R-R Field Background
\vskip.3cm
}
\vskip .5in
{\large
Chen-Te Ma$^{a, b, c, d, e}$ \footnote{e-mail address: yefgst@gmail.com}
\\
\vskip 1mm
}
{\sl
$^a$
Asia Pacific Center for Theoretical Physics,\\
Pohang University of Science and Technology, 
Pohang 37673, Gyeongsangbuk-do, South Korea. 
\\
$^b$
Guangdong Provincial Key Laboratory of Nuclear Science,\\ 
Institute of Quantum Matter,
South China Normal University, Guangzhou 510006, Guangdong, China.
\\
$^c$
School of Physics and Telecommunication Engineering,\\
 South China Normal University, 
 Guangzhou 510006, Guangdong, China.
\\
$^d$
Guangdong-Hong Kong Joint Laboratory of Quantum Matter,\\
 Southern Nuclear Science Computing Center, 
South China Normal University, Guangzhou 510006, Guangdong, China. 
\\
$^e$ 
The Laboratory for Quantum Gravity and Strings,\\
 Department of Mathematics and Applied Mathematics,
University of Cape Town, Private Bag, Rondebosch 7700, South Africa.
}
\\
\vskip 1mm
\vspace{40pt}
\end{center}

\newpage
\begin{abstract}
The volume-preserving diffeomorphism is a key feature that characterizes the large constant R-R ($p-1$)-form field background in a D$p$-brane theory. 
It represents a symmetry of the theory that preserves the volume of space. 
To describe this symmetry, we introduce the concept of the ($p-1$)-bracket, which generates the volume-preserving diffeomorphism. 
The ($p-1$)-bracket is a mathematical operation that acts on ($p-1$)-forms and encodes the transformation of the background field under the symmetry. 
To generalize the ($p-1$)-bracket, we can apply it to the non-Abelian one-form gauge field, which is relevant in gauge theories with non-Abelian gauge groups. 
This allows us to extend the concept of volume-preserving diffeomorphism and its associated symmetry to non-Abelian gauge theories. 
When considering D-branes and T-duality, we introduce the transverse coordinates of the branes. 
T-duality is a symmetry transformation that relates String Theory compactified on different backgrounds. 
It exchanges the momentum and winding modes of strings and leads to an equivalence between theories with different numbers of dimensions. 
By incorporating T-duality and the generalized bracket, a general expression for the action in D$p$-branes can be derived when $p\le 6$. 
This result connects the existing construction of D$p$-branes with our generalized bracket, illustrating the relationship between the symmetry and its associated transformations and the dynamics of the branes. 
In addition, we can discuss the non-Abelianization of the ($p-2$)-form gauge potential. 
This process involves generalizing the concept of non-Abelian gauge fields to higher-form gauge potentials. 
By extending the Lagrangian description of a single D-brane to multiple D-branes, a similar Lagrangian description can be established for both cases, highlighting the common underlying structure and symmetry properties. 
Our developments demonstrate the interplay between symmetries, gauge fields, and D-brane dynamics, providing a deeper understanding of the underlying principles within D-branes.
\end{abstract}
\end{titlepage}

\section{Introduction}
\label{sec:1}
\noindent
In String Theory, open strings have endpoints that can be attached to certain surfaces or hypersurfaces called Dirichlet (D)-branes. 
The endpoints of the open string lie on a ($p+1$)-dimensional hypersurface, where $p$ represents the number of spatial dimensions of the D-brane. 
The positions of the endpoints on a D-brane induce non-commutativity in the target spaces.  
The Seiberg-Witten (SW) map is a transformation that relates the commutative and non-commutative descriptions of certain physical quantities in String Theory \cite{Seiberg:1999vs,Cornalba:1999ah,Okawa:1999cm,Asakawa:1999cu,Ishibashi:1999vi}. 
In the non-commutative description, the ordinary product of functions on the target space is replaced by the Moyal product. 
The non-commutative description, which incorporates the Moyal product, includes effects of stringy corrections represented by the parameter $\alpha^{\prime}$, known as the inverse string tension. 
This parameter appears in the non-commutative theory as the inverse of the Neveu-Schwarz–Neveu-Schwarz (NS-NS) two-form field background. 
The presence of this background field leads to the non-commutativity of the target space coordinates and introduces the stringy corrections. 
\\

\noindent
The non-commutative geometry is also applicable in M-branes. 
The entropy of coincident $N$ D-branes scales as $N^2$. 
However, M5-branes (specifically, M2 branes ending on M5 branes or M2-M5 branes) scale as $N^3$. 
The scaling difference in entropy suggests that the gauge formulation of M-branes (specifically M5-branes) should be novel compared to D-branes \cite{Chu:2011fd,Chu:2012um}. 
The stack of M2-branes provides the Nambu-Poisson (NP) M5-brane theory, which describes the M5-brane in a large three-form field background (referred to as $C$-field) \cite{Ho:2008nn, Ho:2008ei}. 
NP M5-brane theory introduces a new gauge symmetry known as {\it volume-preserving diffeomorphism} (VPD) \cite{Ho:2008nn, Ho:2008ei}. 
The Nambu-Poisson bracket (or three-bracket) generates this symmetry. 
The double dimensional reduction in the direction of the three-form field background yields a D4-brane in a large NS-NS $B$-field background \cite{Ho:2008ve}. 
The non-commutative description of a D4-brane (D2-D4 brane) in a large Ramond-Ramond (R-R) $C$-field background appears through the double dimensional reduction, not alone the direction of $C$-field background \cite{Ho:2011yr}. 
Conditions are presented to describe the unique bosonic sector of a D$p$-brane (specifically D($p-2$)-D$p$ brane) in a large R-R ($p-1$)-form field background \cite{Ho:2013paa}. 
These conditions involve partial Lorentz symmetry, gauge symmetry, field content, duality relations, and the behavior in the large R-R field background \cite{Ho:2013paa}:
\begin{itemize}
\item{Partial Lorentz Symmetry: SO(1, 1)$\times$SO($p-1$)$\times$SO($9-p$); 
}
\item{Gauge Symmetry: U(1) and VPD;}
\item{Field Content: one-form gauge field, non-dynamical $(p-2)$-form gauge potential, and transversal scalar fields;} 
\item{Duality: D$p$-brane is relevant to D($p\pm 1$)-brane via Target-Space duality (T-duality);} 
\item{Leading Order in Large R-R Field Background: the low-energy theory agrees with the trivial background but with a different metric. 
}
\end{itemize}
The scaling limit for the R-R D$p$-brane provides a good approximation \cite{Ho:2013paa}:
\bea
l_s\sim\epsilon^{1/2}; \ 
g_s\sim\epsilon^{-1/2}; \
C_{\dot{\mu_1}\dot{\mu_2}\cdots\dot{\mu}_{p-1}}\sim\epsilon^0; \  
g_{\alpha\beta}\sim\epsilon^0; \ 
g_{\dot{\mu}\dot{\nu}}\sim\epsilon; \ 
\epsilon\rightarrow 0,
\eea
wich involves the string length $l_s\equiv(\alpha^{\prime})^{1/2}$, string coupling constant $g_s$, spacetime metrics $g_{\alpha\beta}$ and $g_{\dot{\mu}\dot{\nu}}$, and the constant R-R ($p-1$)-form field background $C_{\dot{\mu}_1\dot{\mu}_2\cdots\dot{\mu}_{p-1}}$. 
Therefore, the field background only has one non-trivial component.  
The  $\alpha=0, 1$ denotes the directions of spacetime not parallel to the field background. 
We denote other directions from $\dot{\mu}=2, 3, \cdots, p$. 
The worldvolume coordinates are $x^{\alpha}$ and $y^{\dot{\mu}}$. 
The {\it ($p-1$)-bracket} generates VPD and the large ($p-1$)-form field background \cite{Ho:2013paa}. 
The study emphasizes that the ($p-1$)-bracket strongly constrains the R-R D-brane, indicating the significance of the non-commutative description in understanding their properties. 
\\

\noindent
In this paper, we aim to generalize the ($p-1$)-bracket to multiple D$p$-branes and establish its connection to the non-commutative description of D-branes in a large R-R field background. 
The ($p-1$)-bracket is defined for a single D-brane and is given by 
\bea
\{f_1, f_2 ,\cdots, f_{p-1}\}_{(p-1)}\equiv \epsilon^{\dot{\mu}_1\dot{\mu}_2\cdots\dot{\mu}_{p-1}}(\partial_{\dot{\mu}_1}f_1)(\partial_{\dot{\mu}_2f_2})\cdots(\partial_{\dot{\mu}_{p-1}}f_{p-1}).
\eea 
It satisfies a generalized Jacobi identity, which can be written as 
\bea
&&
\{f_1, f_2, \cdots, f_{p-2}, \{g_1, g_2, \cdots, g_{p-1}\}_{(p-1)}\}_{(p-1)}
\nn\\
&=&\{\{ f_1, f_2, \cdots, f_{p-2}, g_1\}_{(p-1)}, \cdots, g_{p-1}\}_{(p-1)}
\nn\\
&&
+\{g_1, \{f_1, f_2, \cdots, f_{p-2}, g_2\}_{(p-1)}, \cdots, g_{p-1}\}_{(p-1)}
+\cdots
\nn\\
&&
+\{ g_1, g_2, \cdots, g_{p-2}, \{f_1, f_2, \cdots, f_{p-2}, g_{p-1}\}_{(p-1)}\}_{(p-1)}.  
\eea 
This identity expresses the compatibility of the ($p-1$)-bracket with the bracket operation itself. 
We then consider the ($p-1$)-bracket of ($p-1$) VPD covariant fields $F_j$. 
The VPD transformation of $F_j$ is given by  
\bea
\hat{\delta}_{\hat{\Lambda}} F_j=\{f_1, f_2, \cdots, f_{p-2}, F_j\}_{(p-1)}=\hat{\kappa}^{\dot{\mu}}\partial_{\dot{\mu}}F_j, 
\eea
where $\hat{\delta}_{\hat{\Lambda}}$ denotes the gauge transformation, and $\hat{\kappa}^{\dot{\mu}}$ is defined as 
\bea  
\hat{\kappa}^{\dot{\mu}}=\epsilon^{\dot{\mu}_1\dot{\mu}_2\cdots\dot{\mu}_{p-2}\dot{\mu}}
(\partial_{\dot{\mu}_1}f_1)(\partial_{\dot{\mu}_2}f_2)\cdots (\partial_{\dot{\mu}_{p-2}}f_{p-2}). 
\eea 
The ($p-1$)-bracket of these VPD covariant fields is also VPD covariant, i.e., it transforms as 
\bea
\hat{\delta}_{\hat{\Lambda}}\{F_1, F_2, \cdots, F_{p-1}\}_{(p-1)}
=\hat{\kappa}^{\dot{\mu}}\partial_{\dot{\mu}}\{F_1, F_2, \cdots, F_{p-1}\}_{(p-1)}. 
\eea
Furthermore, the vector $\hat{\kappa}^{\dot{\mu}}$ generating the VPD transformations is divergenceless $\partial_{\dot{\mu}}\hat{\kappa}^{\dot{\mu}}=0$. 
These properties ensure that the bracket exhibits a manifest VPD symmetry structure. 
\\

\noindent
The Ref. \cite{Ho:2011yr} proposed the non-commutative multiple D$p$-branes in a large R-R ($p-1$)-form field background. 
The field contents of this theory include a one-form U($N$) gauge field $\hat{a}$ and a non-dynamical ($p-2$)-form U(1) gauge potential $\hat{b}$ \cite{Ho:2011yr}. 
By integrating out the non-dynamical gauge potential $\hat{b}$, one can obtain the Yang-Mills gauge theory \cite{Ho:2011yr}. 
The construction of this theory is non-trivial due to the presence of the covariant field strengths. 
The field strengths are defined to incorporate the non-commutative nature of the D-branes and the R-R field background:  
\bea
\hat{{\cal H}}^{\dot{\mu}_1\dot{\mu_2}\cdots\dot{\mu}_{p-1}}&\equiv& g^{p-2}
\{\hat{X}^{\dot{\mu}_1}, \hat{X}^{\dot{\mu}_2}, \cdots, \hat{X}^{\dot{\mu}_{p-1}}\}_{(p-1)}-\frac{1}{g}\epsilon^{\dot{\mu}_1\dot{\mu}_2\cdots\dot{\mu}_{p-1}};
\nn\\
\hat{{\cal F}}_{\dot{\mu}\dot{\nu}}&\equiv& \hat{F}_{\dot{\mu}\dot{\nu}}
+g(\partial_{\dot{\sigma}}\hat{b}^{\dot{\sigma}}\hat{F}_{\dot{\mu}\dot{\nu}}
-\partial_{\dot{\mu}}\hat{b}^{\dot{\sigma}}\hat{F}_{\dot{\sigma}\dot{\nu}}
-\partial_{\dot{\nu}}\hat{b}^{\dot{\sigma}}\hat{F}_{\dot{\mu}\dot{\sigma}});
\nn\\
\hat{{\cal F}}_{\alpha\dot{\mu}}&\equiv& (\hat{V}^{-1})_{\dot{\mu}}{}^{\dot{\nu}}(\hat{F}_{\alpha\dot{\nu}}+g\hat{F}_{\dot{\nu}\dot{\delta}}\hat{B}_{\alpha}{}^{\dot{\delta}});
\nn\\
\hat{{\cal F}}_{\alpha\beta}&\equiv&\hat{F}_{\alpha\beta}+
g(-\hat{F}_{\alpha\dot{\mu}}\hat{B}_{\beta}{}^{\dot{\mu}}-\hat{F}_{\dot{\mu}\beta}\hat{B}_{\alpha}{}^{\dot{\mu}})+
g^2\hat{F}_{\dot{\mu}\dot{\nu}}\hat{B}_{\alpha}{}^{\dot{\mu}}\hat{B}_{\beta}{}^{\dot{\nu}}. 
\eea
Here, $\hat{V}_{\dot{\nu}}{}^{\dot{\mu}}$ and $\hat{X}^{\dot{\mu}}$ are defined as follows: 
\bea
\hat{V}_{\dot{\nu}}{}^{\dot{\mu}}\equiv\delta_{\dot{\nu}}{}^{\dot{\mu}}+g\partial_{\dot{\nu}}\hat{b}^{\dot{\mu}}; \qquad 
\hat{X}^{\dot{\mu}}\equiv\frac{y^{\dot{\mu}}}{g}+\hat{b}^{\dot{\mu}}.   
\eea 
These field strengths and the field $\hat{X}^{\dot{\mu}}$ are covariant under the closed gauge transformation. 
The gauge transformation rules for the fields are given by:
\bea
\hat{\delta}_{\hat{\Lambda}}\hat{b}^{\dot{\mu}}&=&\hat{\kappa}^{\dot{\mu}}
+g\hat{\kappa}^{\dot{\nu}}\partial_{\dot{\nu}}\hat{b}^{\dot{\mu}};
\nn\\
\hat{\delta}_{\hat{\Lambda}}\hat{a}_{\dot{\mu}}&=&\partial_{\dot{\mu}}\hat{\lambda}+
i\lbrack\hat{\lambda}, \hat{a}_{\dot{\mu}}\rbrack
+g(\hat{\kappa}^{\dot{\nu}}\partial_{\dot{\nu}}\hat{a}_{\dot{\mu}}
+\hat{a}_{\dot{\nu}}\partial_{\dot{\mu}}\hat{\kappa}^{\dot{\nu}});
\nn\\
\hat{\delta}_{\hat{\Lambda}}\hat{a}_{\alpha}&=&\partial_{\alpha}\hat{\lambda}+
i\lbrack\hat{\lambda}, \hat{a}_{\alpha}\rbrack
+g(\hat{\kappa}^{\dot{\nu}}\partial_{\dot{\nu}}\hat{a}_{\alpha}
+\hat{a}_{\dot{\nu}}\partial_{\alpha}\hat{\kappa}^{\dot{\nu}}), 
\eea 
where the commutator is $\lbrack{\cal O}_1, {\cal O}_2\rbrack\equiv {\cal O}_1^c{\cal O}_2^d(T^cT^d-T^dT^c)$. 
The $T^c$ is the generator of a Lie algebra. 
We denote the indices of a Lie algebra by $c, d$. 
The field strength ${\cal F}_{\dot{\mu}\dot{\nu}}$ needs to be modified when $p>4$ due to the T-duality \cite{Ho:2013paa}. 
In the case of a single D-brane, the field strength is given by \cite{Ho:2013paa}
\bea
\hat{{\cal F}}_{\dot{\mu}\dot{\nu}}=\frac{g^{p-3}}{(p-3)!}\epsilon_{\dot{\mu}\dot{\nu}\dot{\mu}_1\cdots\dot{\mu}_{p-3}}
\{\hat{X}^{\dot{\mu}_1}, \cdots, \hat{X}^{\dot{\mu}_{p-3}}, \hat{a}_{\dot{\rho}}, \hat{y}^{\dot{\rho}}\}.   
\eea 
It is convenient when transforming the ($p-2$)-form field, $\hat{b}_{\dot{\mu}_2\dot{\mu}_3\cdots\dot{\mu}_{p-1}}$ to the one-form field
\bea
\hat{b}^{\dot{\mu}_1}\equiv\frac{1}{(p-2)!}\epsilon^{\dot{\mu}_1\dot{\mu}_2\cdots\dot{\mu}_{p-1}}\hat{b}_{\dot{\mu}_2\dot{\mu}_3\cdots\dot{\mu}_{p-1}}.  
\eea 
For example, the field $\hat{X}^{\dot{\mu}}$ is VPD covariant. 
The coupling constant $g$ is defined as the inverse of the ($p-1$)-form field background $g\equiv 1/C_{23\cdots p}$. 
Spacetime indices are raised or lowered by using the flat metric, $\eta_{AB}\equiv\mathrm{diag}(-, +, +, \cdots, +)$, 
where $A\equiv(\alpha, \dot{\mu})$. 
The fields, $\hat{b}^{\dot{\mu}}$, $\hat{B}_{\alpha}{}^{\dot{\mu}}$, and $\hat{\kappa}^{\dot{\mu}}$, take the value of U(1), and 
\bea
\hat{F}_{AB}\equiv\partial_A\hat{a}_B-\partial_B\hat{a}_A-i\lbrack \hat{a}_A, \hat{a}_B\rbrack
\eea
is the ordinary covariant field strength (for the trivial background or $g=0$). 
The $\hat{B}_{\alpha}{}^{\dot{\mu}}$ satisfies the following equation 
\bea
\hat{V}_{\dot{\mu}}{}^{\dot{\nu}}(\partial^{\alpha}\hat{b}_{\dot{\nu}}-\hat{V}^{\dot{\rho}}{}_{\dot{\nu}}\hat{B}^{\alpha}{}_{\dot{\rho}})
+\epsilon^{\alpha\beta}\hat{F}^{\mathrm{U}(1)}_{\beta\dot{\mu}}
+g\epsilon^{\alpha\beta}\hat{F}^{\mathrm{U}(1)}_{\dot{\mu}\dot{\nu}}\hat{B}_{\beta}{}^{\dot{\nu}}=0,
\eea
 where $\hat{F}_{AB}^{\mathrm{U}(1)}\equiv\partial_A\hat{a}^{\mathrm{U}(1)}_B-\partial_B\hat{a}^{\mathrm{U}(1)}_A$ is the Abelian strength. 
The gauge potential $\hat{a}_{\alpha}$ cannot appear in the R-R D4-brane from the double dimensional reduction of the NP M5-brane directly \cite{Ho:2011yr}. 
It is necessary to introduce the Lagrange multiplier to extract $\hat{a}_{\alpha}$ through the dualing $\epsilon^{\dot{\mu}\dot{\nu}\dot{\lambda}}\partial_{\dot{\nu}}\hat{b}_{\alpha\dot{\lambda}}$ \cite{Ho:2011yr}. 
After integrating the Lagrangian multiplier, the $\hat{B}_{\alpha}{}^{\dot{\mu}}$ appears in the dual action \cite{Ho:2011yr}. 
 \\

\begin{figure}
\begin{center}
\includegraphics[width=1.\textwidth]{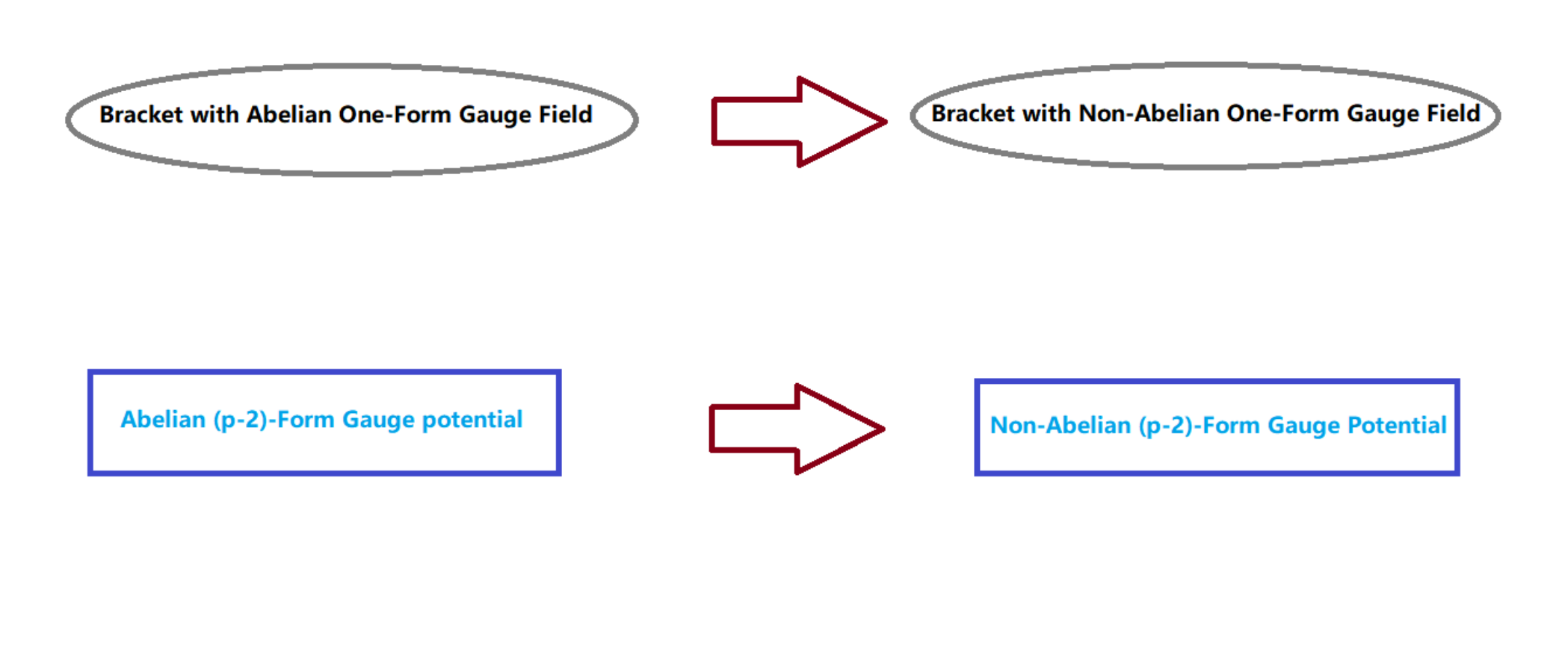}
\end{center}
\caption{Our results show the generalization of the bracket and the ($p-2$)-form gauge potential.}
\label{Generalization.pdf}
\end{figure}

\noindent
According to Ref. \cite{Ma:2020msx}, the author applied the SW map to the construction and then extended it to the U($N$) gauge group and stated that the result is not compatible with the findings of Ref. \cite{Ho:2011yr}. 
The issue arises because the ($p-2$)-form gauge potential is Abelian \cite{Ma:2020msx}. 
To address this problem, we propose a generalization of the ($p-1$)-bracket to reproduce the results of Ref. \cite{Ho:2011yr}. 
Furthermore, we introduce a Lagrangian description through the non-Abelianization of the ($p-2$)-form gauge potential. 
By doing so, the issue disappears, as we establish a universal description between the Abelian and non-Abelian sectors. 
The main results of our work are presented in Fig. \ref{Generalization.pdf}. 
In summary, we make the following key points:
\begin{itemize}
\item{We generalize the ($p-1$)-bracket construction from a single D$p$-brane to multiple D$p$-branes by replacing $\partial$ with the ordinary covariant derivative ($D$). 
The generalization of the bracket yields the generation of VPD and provides strong constraints on the D-branes theory. 
}
\item{We perform dimensional reduction for the bracket recursively, starting from D9-branes. 
When $p\le 6$, we find compact expressions for the interaction terms relevant to the brackets. 
When the ($p-2$)-form gauge potential takes U(1) values, it aligns with the existing formulation. 
However, for the U($N$) case, the interaction between the bracket and the commutator becomes more complex but should be similar. 
}
\item{We maintain U(1) symmetry group for the VPD gauge parameter while introducing a non-Abelian $\hat{b}$. 
This implies that the R-R field background is only present in the U(1) sector. 
Non-Abelianization involves replacing $\partial$ with $D$, allowing us to write a Lagrangian for the non-Abelian ($p-2$)-form gauge potential. 
We discuss the explicit definitions of the field strengths, taking into account  various terms involving the coordinates, gauge fields, and the R-R field background. 
By integrating the non-dynamical gauge potential up to the leading order, we show that the resulting theory can be described by a Yang-Mills gauge theory. 
The Lagrangian description of multiple D-branes resembles that of a single D-brane in this approach.  
}
\end{itemize} 

\noindent 
The organization of this paper is as follows: We generalize the ($p-1$)-bracket in Sec.~\ref{sec:2}. 
We then show the compact expression for interacting terms from this bracket when $p\le 6$ in Sec.~\ref{sec:3}. 
The result of the non-Abelianization ($p-2$)-form gauge potential is  in Sec.~\ref{sec:4}.   
Finally, we discuss our results and conclude in Sec.~\ref{sec:5}.

\section{($p-1$)-Bracket}
\label{sec:2}
\noindent
We first study a single-component $\hat{b}^{\dot{\mu}}$ for reproducing the result of Ref. \cite{Ho:2011yr}. 
The calculation is more simple than using U($N$) $\hat{b}^{\dot{\mu}}$. 
The T-duality induces the bracket and commutator in the D-branes. 
Because the U($N$) $\hat{b}$ and $\hat{X}^I$ have a similar coupling with $\hat{a}_{\dot{\mu}}$, the commutator $\lbrack\hat{X}^{\dot{\mu}}, \hat{X}^{I}\rbrack$ should appear in the R-R D-branes. 
The similar coupling structure is enough to learn the non-Abelian $\hat{b}$ result from the U(1) case. 
The non-Abelian generalization of $\hat{b}^{\dot{\mu}}$ should provide similar but more complicated interacting terms to the Lagrangian. 
\\

\noindent
The VPD covariant and U(1) invariant object is 
\bea
\hat{{\cal O}}_{nmml}\equiv\bigg\{\hat{X}^{\dot{\mu}_1}, \cdots, \hat{X}^{\dot{\mu}_n}, 
\hat{a}_{\dot{\nu}_1}, \cdots, \hat{a}_{\dot{\nu}_m}, 
\frac{y^{\dot{\nu}_1}}{g}, \cdots, \frac{y^{\dot{\nu}_m}}{g}, 
\hat{X}^{I_1}, \cdots, \hat{X}^{I_l}\bigg\}_{(p-1)}, 
\eea
where 
\bea
n, m, l\ge 0;\ n+2m+l=p-1,  
\eea
in the R-R D$p$-brane \cite{Ho:2013paa}. 
The D$p$-brane Lagrangian relevant to this ($p-1$)-bracket is \cite{Ho:2013paa}
\bea
{\cal L}_0=-\frac{g^{2(p-2)}}{2}\sum_{n, m, l\in S}C^{p-1}_{nmml}({\cal O}_{nmml})^2, 
\eea
where the indices ($n, m, l$) are in the set  
\bea
S\equiv\{(n, m, l)| n, m, l\ge 0;\ n+2m+l=p-1\},  
\eea
and the coefficient is 
\bea
C^{p-1}_{nmml}\equiv\frac{1}{n!m!m!l!}. 
\eea
The ${\cal L}_0$ is closed for the T-duality (or dimensional reduction) \cite{Ho:2013paa}. 
Therefore, the symmetry structure from this $(p-1)$-bracket restricts the dynamics of the D-brane.  
\\ 

\noindent 
Now we demonstrate the generalization of the U($N$) symmetry group in D4-branes: 
\begin{itemize}
\item{
Covariant Field Strength: 
\bea
\hat{{\cal F}}_{\dot{\mu}\dot{\nu}}=g^2\epsilon_{\dot{\mu}\dot{\nu}\dot{\rho}}
\bigg\{\hat{X}^{\dot{\rho}}, \hat{a}_{\dot{\sigma}}, \frac{y^{\dot{\sigma}}}{g}\bigg\}_{(3)}; 
\eea
}
\item{Covariant Derivative of Scalar Fields: 
\bea
\hat{D}_{\dot{\mu}}\hat{X}^I=\frac{g^2}{2}\epsilon_{\dot{\mu}\dot{\nu}\dot{\rho}}
\{\hat{X}^{\dot{\nu}}, \hat{X}^{\dot{\rho}}, \hat{X}^I\}_{(3)},  
\eea
where the gauge transformation of scalar fields is 
\bea
\hat{\delta}_{\hat{\Lambda}}\hat{X}^I=i\lbrack\hat{\lambda}, \hat{X}^I\rbrack+g\hat{\kappa}^{\dot{\rho}}\partial_{\dot{\rho}}\hat{X}^I, 
\eea
derived from the dimensional reduction on the $p$-th direction (or the following replacement):
\bea
\hat{a}_p\rightarrow \hat{X}^{I=p}; \qquad \partial_p(\mathrm{field})\rightarrow 0.  
\eea
}
\item{Commutator of Scalar Fields:
\bea
\lbrack\hat{X}^I, \hat{X}^J\rbrack\Rightarrow\frac{g^3}{3!}\epsilon_{\dot{\mu}\dot{\nu}\dot{\rho}}
\{\hat{X}^{\dot{\mu}}, \hat{X}^{\dot{\nu}}, \hat{X}^{\dot{\rho}}\}_{(3)}
\lbrack\hat{X}^I, \hat{X}^I\rbrack,  
\eea
where $\Rightarrow$ means the equivalence up to the leading order in $g$. 
}
\end{itemize}
The NP-bracket introduces the necessary terms in multiple D4-branes. 
We change the ordinary derivative to another derivative operator ${\cal D}$ in the bracket: 
\bea
{\cal D}_{\dot{\mu}}\hat{X}^I\equiv \partial_{\dot{\mu}}\hat{X}^I-i\lbrack\hat{a}_{\dot{\mu}}, \hat{X}^I\rbrack\equiv D_{\dot{\mu}}X^I; 
\qquad 
{\cal D}_{\dot{\mu}}\hat{a}_{\dot{\nu}}\equiv(\partial_{\dot{\mu}}-i\hat{a}_{\dot{\mu}})\hat{a}_{\dot{\nu}},  
\eea 
where $D_{\dot{\mu}}$ is the ordinary covariant derivative. 
For the U(1) field, the ${\cal D}_{\dot{\mu}}$ is the same as the ordinary derivative $\partial_{\dot{\mu}}$. 
The combination of $a_{\dot{\mu}}$ and $y^{\dot{\mu}}$ shows the ordinary covariant field strength 
\bea
({\cal D}_{\lbrack\dot{\mu}}\hat{a}_{\dot{\rho}})(\partial_{\dot{\nu}\rbrack}y^{\dot{\rho}})=\hat{F}_{\dot{\mu}\dot{\nu}},  
\eea
in which we use the notation 
\bea
A_{\lbrack\dot{\mu}}B_{\dot{\nu}\rbrack}\equiv A_{\dot{\mu}}B_{\dot{\nu}}-A_{\dot{\nu}}B_{\dot{\mu}}. 
\eea 
\\ 

\noindent 
We can show that ${\cal O}_{nmml}$ is the VPD- and U($N$)-covariant object when promoting the symmetry group of $\hat{a}_{\dot{\mu}}$ from U(1) to U($N$). 
The $\hat{X}^{\dot{\mu}}$ and $\hat{X}^I$ are VPD covariant. 
Because $y^{\dot{\mu}}$ is not a dynamical field, the VPD transformation cannot act on it. 
The $\hat{a}_{\dot{\mu}}$ is not VPD covariant. 
We need to check the combination of $\hat{a}_{\dot{\mu}}$ and $y^{\dot{\mu}}$, 
\bea
\hat{\delta}_{\hat{\Lambda}}\big(({\cal D}_{\lbrack\dot{\mu}}\hat{a}_{\dot{\rho}})(\partial_{\dot{\nu}\rbrack}y^{\dot{\rho}})\big)
=-i\lbrack \hat{F}_{\dot{\mu}\dot{\nu}}, \hat{\lambda}\rbrack
+g\hat{\kappa}^{\dot{\rho}}\partial_{\dot{\rho}}\hat{F}_{\dot{\mu}\dot{\nu}}
+g\big(({\cal D}_{\dot{\mu}}\hat{a}_{\dot{\rho}})(\partial_{\dot{\nu}}\hat{\kappa}^{\dot{\rho}})
-({\cal D}_{\dot{\nu}}\hat{a}_{\dot{\rho}})(\partial_{\dot{\mu}}\hat{\kappa}^{\dot{\rho}})\big). 
\eea 
The first term of the right-hand side shows the U($N$) covariant. 
The second term shows the VPD covariant. 
The third term shows the non-VPD covariant, but the VPD covariant $y^{\dot{\mu}}$:  
\bea
\hat{\delta}_{\hat{\Lambda}}y^{\dot{\mu}}=g\hat{\kappa}^{\dot{\nu}}\partial_{\dot{\nu}}y^{\dot{\mu}}=g\hat{\kappa}^{\dot{\mu}}
\eea  
can generate this term. 
The gauge transformation of the pair $(\hat{\delta}_{\hat{\Lambda}}\hat{a}_{\dot{\mu}}, y^{\dot{\mu}})$ is equivalent to treating the pair as a VPD covariant object. 
A similar result already happens in R-R D-brane \cite{Ho:2013paa}. 
The non-VPD covariant part is due to the ordinary derivative $\partial_{\dot{\mu}}$. 
Therefore, we can apply the generalized Jacobi identity to show the VPD covariance. 
The $\hat{F}_{\dot{\mu}\dot{\nu}}$ and $D_{\dot{\mu}}$ are U($N$) covariant. 
Hence the $(p-1)$-bracket is VPD- and U($N$)-covariant. 
The non-Abelian generalization of this bracket is not closed under the T-duality. 
It is also necessary to introduce the commutator. 
Later we will recursively use the dimensional reduction to show the interaction between the bracket and commutator. 
 
\section{T-Duality}
\label{sec:3}
\noindent 
Now we perform the T-duality for the single-component $b^{\dot{\mu}}$ case. 
The infinitely recursive operation of dimensional reduction should introduce infinite interaction terms. 
Therefore, it is hard to write the analytical expression for the action. 
We begin with D9-branes or the pure gauge theory. 
The field contents relevant to the 8-form bracket are $\hat{a}_{\dot{\mu}}$ and $\hat{b}^{\dot{\mu}}$. 
The Lagrangian relevant to the 8-form bracket is 
\bea
&&
{\cal L}_1^{(p=9)}
\nn\\
&=&
-\frac{g^{2(p-2)}}{2}\sum_{n, m\in S_0}C^{p-1}_{nmm0}
\mathrm{Str}\bigg\lbrack\bigg(\bigg\{\hat{X}^{\dot{\mu}_1}, \cdots, \hat{X}^{\dot{\mu}_n}, 
\hat{a}_{\dot{\nu}_1}, \cdots, \hat{a}_{\dot{\nu}_m}, 
\frac{y^{\dot{\nu}_1}}{g}, \cdots, \frac{y^{\dot{\nu}_m}}{g}\bigg\}_{(p-1)}\bigg)^2\bigg\rbrack, 
\nn\\
\eea 
where
\bea
\mathrm{Str}({\cal O}_1{\cal O}_2\cdots{\cal O}_n)&\equiv&
\mathrm{Tr}\big(\mathrm{Sym}({\cal O}_1{\cal O}_2\cdots{\cal O}_n)\big); 
\nn\\
\mathrm{Sym}({\cal O}_1{\cal O}_2\cdots{\cal O}_n)&\equiv&\frac{1}{n!}({\cal O}_1{\cal O}_2\cdots {\cal O}_n+ \mathrm{all\ permutations}).
\eea 
The objects ${\cal O}_j$ indicate the fields $\hat{X}^{\dot{\mu}}$, $\hat{a}_{\dot{\mu}}$, and the coordinates $y^{\dot{\mu}}$. 
When considering the lower dimensional worldvolume theory, the ${\cal O}_j$ also indicates the commutator, $\lbrack\hat{X}^{I_1}, \hat{X}^{I_2}\rbrack$.  
The indices ($n, m$) are in the set  
\bea
S_0\equiv\{(n, m)| n, m\ge 0;\ n+2m=p-1\}.  
\eea
When performing the dimensional reduction in the $p$-th direction, we fix the following gauge condition
\bea
\hat{b}^{\dot{\mu}=p}=0
\eea
for simplifying the expression of the action. 
In summary, we derive the D($p-1$)-branes from D$p$-branes by the replacement: 
\bea
\hat{a}_{\dot{\mu}=p}\rightarrow\hat{X}^{I=p}; \ 
\hat{b}^{\dot{\mu}=p}\rightarrow 0; \ 
\partial_{\dot{\mu}=p}(\mathrm{field})\rightarrow 0. 
\eea
When $p\le 8$, the ${\cal L}_1^{(p)}$ is 
\bea
&&
{\cal L}_1^{(p)}
\nn\\
&=&
-\frac{g^{2(p-2)}}{2}\sum_{n, m, l\in S_1}C^{p-1}_{nmml}
\nn\\
&&\times
\mathrm{Str}\bigg\lbrack\bigg(\{ \hat{X}^{\dot{\mu}_1}, \cdots, \hat{X}^{\dot{\mu}_{n}}, 
\hat{a}_{\dot{\nu}_1}, \cdots, \hat{a}_{\dot{\nu}_m}, 
\frac{y^{\dot{\nu}_1}}{g}, \cdots, \frac{y^{\dot{\nu}_m}}{g}, 
\hat{X}^{I_l}
\bigg\}_{(p-1)}\bigg)^2\bigg\rbrack,   
\eea
where the indices ($n, m, l$) are in the set  
\bea
S_1\equiv\{(n, m, l)| n, m\ge 0;\ 0\le l \le 9-p;\ n+2m+l=p-1\}.  
\eea
The $l$ is the number of transversal scalar fields. 
Therefore, there are no scalar fields in the brackets when $l=0$. 
We obtain the result of $p=8$ by performing the dimensional reduction in the 9th direction. 
The dimensional reduction identifies $\hat{a}_9$  as $\hat{X}^{I=9}$. 
Therefore, the multiple D8-branes have one more scalar field than $p=9$. 
When $p=8$, we only have one scalar field. 
Hence the Lagrangian does not contain a commutator. 
Later we will show the interaction terms for $p=6, 7$. 
We also use the compact expression to show the field strength, covariant derivative and commutator of scalar fields, and quadratic term of $\hat{{\cal H}}$. 
Our computation shows the result of Ref. \cite{Ho:2011yr}. 

\subsection{$p=7$}
\noindent 
The dimensional reduction in the 7th direction introduces one additional scalar field to D7-branes. 
Two transversal scalar fields can form a bracket. 
Therefore, more terms appear in the Lagrangian when concerning $p=7$. 
When $p\le 7$, we separate the Lagrangian from three terms. 
The first term is  
\bea
&&
{\cal L}_1^{(p)}
\nn\\
&=&
-\frac{g^{2(p-2)}}{2}\sum_{n, m, l\in S_1}C^{p-1}_{nmml}
\nn\\
&&\times
\mathrm{Str}\bigg\lbrack
\bigg(\bigg\{ \hat{X}^{\dot{\mu}_1}, \cdots, \hat{X}^{\dot{\mu}_{n}}, 
\hat{a}_{\dot{\nu}_1}, \cdots, \hat{a}_{\dot{\nu}_m}, 
\frac{y^{\dot{\nu}_1}}{g}, \cdots, \frac{y^{\dot{\nu}_m}}{g}, 
\hat{X}^{I_1}, \cdots, \hat{X}^{I_l}
\bigg\}_{(p-1)}\bigg)^2 \bigg\rbrack.  
\nn\\
\eea
The second term is
\bea
&&
{\cal L}_2^{(p=7)}
\nn\\
&=&
-\frac{g^{2(p-2)+1}}{2}\sum_{n, m\in S_0}(-1)^{m}C^{p-1}_{n(m-1)m0}
\nn\\
&&\times 
\mathrm{Str}\bigg\lbrack
\bigg\{ \hat{X}^{\dot{\mu}_1}, \cdots, \hat{X}^{\dot{\mu}_{n}}, 
\hat{a}_{\dot{\nu}_1}, \cdots, \hat{a}_{\dot{\nu}_{m-1}}, 
\frac{y^{\dot{\nu}_1}}{g}, \cdots, \frac{y^{\dot{\nu}_{m-1}}}{g}, 
\hat{X}^{I_1}, \hat{X}^{I_{2}}
\bigg\}_{(p-1)}
\nn\\
&&\times
\bigg\{ \hat{X}^{\dot{\mu}_1}, \cdots, \hat{X}^{\dot{\mu}_{n}}, 
\hat{a}_{\dot{\bar{\nu}}_1}, \cdots, \hat{a}_{\dot{\bar{\nu}}_{m}}, 
\frac{y^{\dot{\bar{\nu}}_1}}{g}, \cdots, \frac{y^{\dot{\bar{\nu}}_{m}}}{g}
\bigg\}_{(p-1)}
(-i)\lbrack\hat{X}^{I_1}, \hat{X}^{I_2}\rbrack\bigg\rbrack. 
\eea
The last term is 
\bea
&&
{\cal L}_3^{(p=7)}
\nn\\
&=&
\frac{g^{2(p-1)}}{4}\sum_{n, m\in S_0}C^{p-1}_{nmm0}
\nn\\
&&\times
\mathrm{Str}\bigg\lbrack
\bigg(\bigg\{ \hat{X}^{\dot{\mu}_1}, \cdots, \hat{X}^{\dot{\mu}_{n}}, 
\hat{a}_{\dot{\nu}_1}, \cdots, \hat{a}_{\dot{\nu}_m}, 
\frac{y^{\dot{\nu}_1}}{g}, \cdots, \frac{y^{\dot{\nu}_m}}{g}
\bigg\}_{(p-1)}
\lbrack\hat{X}^{I_1}, \hat{X}^{I_2}\rbrack
\bigg)^2\bigg\rbrack.
\eea 
The first term is the straightforward generalization without the commutator. 
The bracket and commutator simultaneously appear in the second and third terms. 
The $l$ in $C^{p-1}_{nmml}$ is the lowest number of traversal scalar fields in the brackets. 
Because the third term also has the square of $(p-1)$-bracket, the dimensional reduction result is similar to the first term (but introduces the additional commutators of scalar fields). 
The second term is the most complicated for performing dimensional reduction. 
A more complicated contraction of indices will appear in the lower dimensions. 
We will show the result for D6-branes. 
Then it is easy to understand the expectation. 

\subsection{$p=6$}
\noindent 
We can generalize ${\cal L}_2^{(p)}$ and ${\cal L}_3^{(p)}$ to include $p=6$: 
\bea
&&
{\cal L}_2^{(p)}
\nn\\
&=&
-\frac{g^{2(p-2)+1}}{2}\sum_{n, m, l\in S_2}(-1)^{m}C^{p-1}_{n(m-1)ml}
\nn\\
&&\times 
\mathrm{Str}\bigg\lbrack
\bigg\{ \hat{X}^{\dot{\mu}_1}, \cdots, \hat{X}^{\dot{\mu}_{n}}, 
\hat{a}_{\dot{\nu}_1}, \cdots, \hat{a}_{\dot{\nu}_{m-1}}, 
\frac{y^{\dot{\nu}_1}}{g}, \cdots, \frac{y^{\dot{\nu}_{m-1}}}{g}, 
\hat{X}^{I_1}, \cdots, \hat{X}^{I_{l+2}}
\bigg\}_{(p-1)}
\nn\\
&&\times
\bigg\{ \hat{X}^{\dot{\mu}_1}, \cdots, \hat{X}^{\dot{\mu}_{n}}, 
\hat{a}_{\dot{\bar{\nu}}_1}, \cdots, \hat{a}_{\dot{\bar{\nu}}_{m}}, 
\frac{y^{\dot{\bar{\nu}}_1}}{g}, \cdots, \frac{y^{\dot{\bar{\nu}}_{m}}}{g}, 
\hat{X}^{I_1}, \cdots, \hat{X}^{I_{l}}
\bigg\}_{(p-1)}
\nn\\
&&\times
(-i)\lbrack\hat{X}^{I_{l+1}}, \hat{X}^{I_{l+2}}\rbrack\bigg\rbrack;  
\eea
\bea
&&
{\cal L}_3^{(p)}
\nn\\
&=&
\frac{g^{2(p-1)}}{4}\sum_{n, m, l\in S_3}C^{p-1}_{nmml}
\nn\\
&&\times
\mathrm{Str}\bigg\lbrack
\bigg(\bigg\{ \hat{X}^{\dot{\mu}_1}, \cdots, \hat{X}^{\dot{\mu}_{n}}, 
\hat{a}_{\dot{\nu}_1}, \cdots, \hat{a}_{\dot{\nu}_m}, 
\frac{y^{\dot{\nu}_1}}{g}, \cdots, \frac{y^{\dot{\nu}_m}}{g}, 
\hat{X}^{\lbrack I_1}, \cdots, \hat{X}^{I_l}
\bigg\}_{(p-1)}
\nn\\
&&\times
\lbrack\hat{X}^{I_{l+1}}, \hat{X}^{I_{l+2}\rbrack}\rbrack
\bigg)^2\bigg\rbrack, .
\eea
where the indices ($n, m, l$) are in the sets: 
\bea
S_2&\equiv&\{(n, m, l)| n\ge 0;\ m=1;\  0\le l\le 7-p;\ n+2m+l=p-1\};   
\nn\\
S_3&\equiv&\{(n, m, l)| n, m\ge 0;\ 0\le l\le 7-p;\ n+2m+l=p-1\}.  
\eea
The ${\cal L}_1^{(p)}$ remains the same expression. 
Each index of traversal scalar fields is not the same in the commutator and bracket. 
Therefore, we introduce the antisymmetrized indices to the transversal scalar fields in the bracket and commutator in ${\cal L}_3^{(p)}$. 
In ${\cal L}_1^{(p)}$ and ${\cal L}_2^{(p)}$, each index of transversal scalar fields is not the same even without the antisymmetrized notation. 
If we proceed with more dimensional reduction, it is easy to find that ${\cal L}_2^{(p)}$ will generate a more complicated contraction between the bracket and the commutator. 
However, the computation is enough to reproduce the result of Ref. \cite{Ho:2011yr}. 
\\

\noindent
When we choose $n=p-3$, $m=1$, and $l=0$ in ${\cal L}_1^{(p)}$, we obtain the quadratic term of $\hat{{\cal F}}_{\dot{\mu}\dot{\nu}}$ 
\bea
-\frac{g^{2(p-2)}}{2(p-3)!}
\mathrm{Str}\bigg\lbrack
\bigg(\bigg\{ \hat{X}^{\dot{\mu}_1}, \cdots, \hat{X}^{\dot{\mu}_{p-3}}, 
\hat{a}_{\dot{\nu}}, \frac{y^{\dot{\nu}}}{g}\bigg\}_{(p-1)}\bigg)^2\bigg\rbrack 
=-\frac{1}{4}\mathrm{Str}\bigg(\hat{{\cal F}}_{\dot{\mu}\dot{\nu}}\hat{{\cal F}}^{\dot{\mu}\dot{\nu}}\bigg),      
\eea 
where 
\bea
\hat{{\cal F}}_{\dot{\mu}\dot{\nu}}=\frac{g^{p-3}}{(p-3)!}\epsilon_{\dot{\mu}\dot{\nu}\dot{\mu}_1\cdots\dot{\mu}_{p-3}}
\{\hat{X}^{\dot{\mu}_1}, \cdots, \hat{X}^{\dot{\mu}_{p-3}}, \hat{a}_{\dot{\rho}}, \hat{y}^{\dot{\rho}}\}.   
\eea 
When $p\le 4$, the gauge sector (without the traversal scalar fields $\hat{X}^I$) relevant to the bracket is the same as in Ref. \cite{Ho:2011yr}. 
The $\hat{{\cal F}}_{\dot{\mu}\dot{\nu}}$ shows a straightforward generalization of a single D-brane (only replaces the U(1) gauge group by U($N$)) when $p>4$. 
The quadratic term of $\hat{{\cal H}}$ also appears in ${\cal L}_1^{(p)}$ when choosing $n=p-1$ and $m=l=0$,
\bea
-\frac{g^{2(p-2)}}{2(p-1)!}
\bigg(\{ \hat{X}^{\dot{\mu}_1}, \cdots, \hat{X}^{\dot{\mu}_{p-1}} 
\}_{(p-1)}\bigg)^2
=-\frac{1}{2(p-1)!}
\bigg(\hat{{\cal H}}_{\dot{\mu_1}\cdots\dot{\mu}_{p-1}}
+\frac{1}{g}\epsilon_{\dot{\mu}_1\cdots\dot{\mu}_{p-1}}\bigg)^2.  
\eea
Because the symmetry group of gauge potential $\hat{b}$ is U(1), the result is the same as in R-R D-brane \cite{Ho:2011yr}. 
Therefore, the result is consistent with Ref. \cite{Ho:2011yr}. 
Now we discuss the transversal scalar fields. 
We first choose $n=p-2$, $m=0$, and $l=1$ in ${\cal L}_1^{(p)}$ to obtain the kinetic term of the scalar fields $\hat{X}^I$, 
\bea
-\frac{g^{2(p-2)}}{2(p-2)!}
\mathrm{Str}\bigg\lbrack
\bigg(\{ \hat{X}^{\dot{\mu}_1}, \cdots, \hat{X}^{\dot{\mu}_{p-2}}, 
\hat{X}^{I}
\}_{(p-1)}\bigg)^2
\bigg\rbrack
=-\frac{1}{2}
\mathrm{Str}\bigg\lbrack
(\hat{D}_{\dot{\mu}}\hat{X}^I)^2\bigg\rbrack,  
\eea
where 
\bea
\hat{D}_{\dot{\mu}}\hat{X}^I
=\frac{(-1)^p}{(p-2)!}g^{p-2}\epsilon_{\dot{\mu}_1\cdots\dot{\mu}_{p-1}}
\{\hat{X}^{\dot{\mu}_2}, \cdots, \hat{X}^{\dot{\mu}_{p-1}}, \hat{X}^I\}. 
\eea 
The commutator of scalar fields also appears in the leading-order term of ${\cal L}_3^{(p)}$ when choosing $n=p-1$ and $m=l=0$ in ${\cal L}_3^{(p)}$,  
\bea
\frac{g^{2(p-1)}}{4(p-1)!}
\mathrm{Str}\bigg\lbrack
\bigg(\{ \hat{X}^{\dot{\mu}_1}, \cdots, \hat{X}^{\dot{\mu}_{p-1}}
\}_{(p-1)}
\lbrack\hat{X}^{I_{1}}, \hat{X}^{I_{2}\rbrack}\rbrack
\bigg)^2
\bigg\rbrack
=
\frac{1}{4}
\mathrm{Str}\big\lbrack
(\lbrack\hat{X}^I, \hat{X}^J\rbrack)^2\big\rbrack+\cdots. 
\eea 
Hence our result also contains the expected terms of transversal scalar fields for the general $p$. 
\\

\noindent
The above calculation keeps the gauge group of $\hat{b}$ as U(1). 
The non-Abelianization needs to concern the consistency from all components of $\hat{{\cal F}}$. 
So far, we did not have such a construction for the R-R D-branes.
Later we will replace the $\partial$ with the $D$. 
We then can show the consistent result from the gauge symmetry.  
The gauge theory also has the same form as in the R-R D-brane. 
Hence we expect that the dimensional reduction should show a similar result (but needs to introduce more commutator terms relevant to $\hat{X}^{\dot{\mu}}$ and $\hat{X}^I$).

\section{Discussion of Non-Abelianization}
\label{sec:4}
\noindent 
The main difficulty for the non-Abelianization is $\hat{B}_{\alpha}{}^{\dot{\mu}}$, which satisfies the non-linear equation in R-R D-brane theory. 
The nonlinearity makes the complexity of the multiple branes with VPD symmetry. 
The study of ($p-1$)-bracket shows the clues. 
The naive generalization (replaces the ordinary derivative with the ordinary covariant derivative) allows the VPD symmetry. 
Our study or the generalization of ($p-1$)-bracket is only for the non-Abelianization of $\hat{a}$. 
Indeed, we can also have a similar generalization for the $\hat{b}$ if the gauge group of the VPD parameter is still U(1). 
Therefore, the R-R field background only lives in the U(1) sector. 
The gauge transformation of $\hat{b}$ is 
\bea
\hat{\delta}_{\hat{\Lambda}}\hat{b}^{\dot{\mu}}
=\hat{\kappa}^{\dot{\mu}}
+i\lbrack\hat{\lambda}, \hat{b}^{\dot{\mu}}\rbrack
+g\hat{\kappa}^{\dot{\nu}}\partial_{\dot{\nu}}\hat{b}^{\dot{\mu}}.
\eea 
The $\hat{X}^{\dot{\mu}}$ is still a VPD- and U($N$)-covariant object. 
Therefore, we can also use the ($p-1$)-bracket to define the covariant field strength 
\bea
\hat{{\cal H}}^{\dot{\mu}_1\dot{\mu_2}\cdots\dot{\mu}_{p-1}}&\equiv& g^{p-2}
\{\hat{X}^{\dot{\mu}_1}, \hat{X}^{\dot{\mu}_2}, \cdots, \hat{X}^{\dot{\mu}_{p-1}}\}_{(p-1)}-\frac{1}{g}\epsilon^{\dot{\mu}_1\dot{\mu}_2\cdots\dot{\mu}_{p-1}}, 
\eea
but now the derivative operator ${\cal D}$ is the ordinary covariant derivative. 
In the R-R D4-brane, the gauge transformation of $\hat{b}$ comes from the NP M5-brane \cite{Ho:2011yr}. 
Therefore, we can use the R-R D4-branes to explore the gauge structure of the M5-brane. 
Our gauge transformation of $\hat{b}$ is consistent with one construction of multiple M5-branes \cite{Chu:2012um}. 
Our realization of VPD symmetry in R-R D-branes is not just a naive guess and has supporting evidence from M2-M5 branes. 
\\

\noindent 
We propose the Lagrangian description of R-R D-branes similar to the single D-brane. 
The Lagrangian for the gauge sector of R-R D$p$-branes theory is  
\bea
{\cal L}_{RR}&=&
{\cal L}_1^{(p)}
+\frac{1}{2}\hat{{\cal F}}_{\alpha\dot{\mu}}\hat{{\cal F}}^{\alpha\dot{\mu}}
+\frac{1}{2g}\epsilon^{\alpha\beta}\hat{{\cal F}}_{\alpha\beta}
\nn\\
&=&
-\frac{1}{2}\hat{{\cal H}}_{\dot{\mu}_1\cdots\dot{\mu}_{p-1}}\hat{{\cal H}}^{\dot{\mu}_1\cdots\dot{\mu}_{p-1}}
-\frac{1}{4}\hat{{\cal F}}_{\dot{\mu}\dot{\nu}}\hat{{\cal F}}^{\dot{\mu}\dot{\nu}}
+\frac{1}{2}\hat{{\cal F}}_{\alpha\dot{\mu}}\hat{{\cal F}}^{\alpha\dot{\mu}}
+\frac{1}{2g}\epsilon^{\alpha\beta}\hat{{\cal F}}_{\alpha\beta}+{\cal O}(g^2),    
\eea 
where ${\cal O}(g^2)$ goes away when $p\le 4$. 
We will use the ($p-1$)-bracket to define $\hat{{\cal H}}_{\dot{\mu}_1\dot{\mu}_2\cdots\dot{\mu}_{p-1}}$ and 
$\hat{{\cal F}}_{\dot{\mu}\dot{\nu}}$. 
Other components of $\hat{{\cal F}}$ need to concern $\hat{B}_{\alpha}{}^{\dot{\mu}}$. 
Therefore, we examine the gauge transformation and modify the non-linear equation, satisfied by $\hat{B}_{\alpha}{}^{\dot{\mu}}$. 
We will first examine the leading-order result. 
Let us remind the result of the R-R D-brane. 
One can do the small $g$ expansion to solve the non-linear equation in R-R D-brane \cite{Ho:2011yr}.  
The $\hat{B}_{\alpha}^{\dot{\mu}}$ is \cite{Ho:2011yr}
\bea
\hat{B}_{\alpha}{}^{\dot{\mu}}=
\partial_{\alpha}\hat{b}^{\dot{\mu}}
+\epsilon_{\alpha\beta}\hat{F}^{\beta\dot{\mu}}+{\cal O}(g). 
\eea
Then one can read the gauge transformation \cite{Ho:2011yr}
\bea
\hat{\delta}_{\hat{\Lambda}}\hat{B}_{\alpha}{}^{\dot{\mu}}=\partial_{\alpha}\hat{\kappa}^{\dot{\mu}}+{\cal O}(g). 
\eea 
Now we replace $\partial$ with the $D$ in the multiple D-branes. 
The non-linear equation becomes
\bea
\hat{B}_{\alpha}{}^{\dot{\mu}}=
D_{\alpha}\hat{b}^{\dot{\mu}}
+\epsilon_{\alpha\beta}\hat{F}^{\beta\dot{\mu}}+{\cal O}(g),  
\eea
where $\hat{F}$ is the ordinary covariant field strength. 
Then we can get the gauge transformation 
\bea
\hat{\delta}_{\hat{\Lambda}}\hat{B}_{\alpha}{}^{\dot{\mu}}=\partial_{\alpha}\hat{\kappa}^{\dot{\mu}}+i\lbrack\hat{\lambda}, \hat{B}_{\alpha}{}^{\dot{\mu}}\rbrack+{\cal O}(g).
\eea 
The non-linear equation that constrains $\hat{B}_{\alpha}{}^{\dot{\mu}}$ is also covariant under the gauge transformation
\bea
i\lbrack\hat{\lambda}, D_{\alpha}\hat{b}^{\dot{\mu}}+\epsilon_{\alpha\beta}\hat{F}^{\beta\dot{\mu}}-\hat{B}_{\alpha}{}^{\dot{\mu}}\rbrack
={\cal O}(g). 
\eea
\\

\noindent 
Now we discuss the low-energy theory in the leading order. 
We first review the R-R D$p$-brane theory \cite{Ho:2011yr}. 
The field strengths are \cite{Ho:2011yr}: 
\bea
\hat{{\cal H}}_{23\cdots p}&=&\hat{H}+{\cal O}(g)\equiv \partial_{\dot{\mu}}\hat{b}^{\dot{\mu}}+{\cal O}(g); 
\nn\\ 
\hat{{\cal F}}_{\dot{\mu}\dot{\nu}}&=&\hat{F}_{\dot{\mu}\dot{\nu}}+{\cal O}(g); 
\nn\\ 
\hat{{\cal F}}_{\alpha\dot{\mu}}&=&\hat{F}_{\alpha\dot{\mu}}+{\cal O}(g);
\nn\\
\frac{1}{2g}\epsilon^{\alpha\beta}\hat{{\cal F}}_{\alpha\beta}&=&
\frac{1}{2g}\epsilon^{\alpha\beta}\hat{F}_{\alpha\beta}
-\epsilon^{\alpha\beta}\hat{F}_{\alpha\dot{\mu}}
(\partial_{\beta}\hat{b}^{\dot{\mu}}+\epsilon_{\beta\gamma}\hat{F}^{\gamma\dot{\mu}})
+{\cal O}(g). 
\eea
The R-R D-brane Lagrangian is \cite{Ho:2011yr} 
\bea 
{\cal L}_{RR}\sim-\frac{1}{2}\hat{H}^2
-\frac{1}{4}\hat{F}_{\dot{\mu}\dot{\nu}}\hat{F}^{\dot{\mu}\dot{\nu}}
-\frac{1}{2}\hat{F}_{\alpha\dot{\mu}}\hat{F}^{\alpha\dot{\mu}}
-\hat{F}_{01}\hat{H}+{\cal O}(g),  
\eea
where $\sim$ is the equivalence up to a total derivative term. 
Because $\hat{b}$ does not have a time-derivative term, we can integrate it, equivalent to substituting \cite{Ho:2011yr}  
\bea
\hat{H}=-\hat{F}_{01}+f, 
\eea
where 
\bea
\partial_{\dot{\mu}}f=0,  
\eea 
to ${\cal L}_{RR}$. 
Therefore, we get the low-energy theory \cite{Ho:2011yr}
\bea
{\cal L}_{RR0}\sim -\frac{1}{4}\hat{F}_{AB}\hat{F}^{AB}-\frac{f^2}{2}+{\cal O}(g). 
\eea
Because $f$ does not interact with the dynamical field, we can ignore it and obtain the expected kinetic term of $\hat{a}_A$. 
The extension of the non-Abelian gauge group is similar. 
We will replace the $\partial$ with $D$ and promote the U(1) invariant quantity to the U($N$) covariant object. 
The field strengths become:
\bea
\hat{{\cal H}}_{23\cdots p}&=&\hat{H}+{\cal O}(g)\equiv D_{\dot{\mu}}\hat{b}^{\dot{\mu}}+{\cal O}(g); 
\nn\\ 
\hat{{\cal F}}_{\dot{\mu}\dot{\nu}}&=&\hat{F}_{\dot{\mu}\dot{\nu}}+{\cal O}(g); 
\nn\\ 
\hat{{\cal F}}_{\alpha\dot{\mu}}&=&\hat{F}_{\alpha\dot{\mu}}+{\cal O}(g);
\nn\\
\frac{1}{2g}\epsilon^{\alpha\beta}\hat{{\cal F}}_{\alpha\beta}&=&
\frac{1}{2g}\epsilon^{\alpha\beta}\hat{F}_{\alpha\beta}
-\epsilon^{\alpha\beta}\hat{F}_{\alpha\dot{\mu}}
(D_{\beta}\hat{b}^{\dot{\mu}}+\epsilon_{\beta\gamma}\hat{F}^{\gamma\dot{\mu}})
+{\cal O}(g). 
\eea 
The Lagrangian description of R-R D-branes remains the same form 
\bea 
{\cal L}_{RR}\sim-\frac{1}{2}\hat{H}^2
-\frac{1}{4}\hat{F}_{\dot{\mu}\dot{\nu}}\hat{F}^{\dot{\mu}\dot{\nu}}
-\frac{1}{2}\hat{F}_{\alpha\dot{\mu}}\hat{F}^{\alpha\dot{\mu}}
-\hat{F}_{01}\hat{H}+{\cal O}(g).
\eea
Integrating out the non-dynamical field $\hat{b}$ is equivalent to substituting 
\bea
\hat{H}=-\hat{F}_{01}+f, 
\eea 
where 
\bea
D_{\dot{\mu}}f=0. 
\eea
The $f$ also does not interact with $\hat{a}_A$, 
\bea
{\cal L}_{RR0}\sim -\frac{1}{4}\hat{F}_{AB}\hat{F}^{AB}-\frac{f^2}{2}+{\cal O}(g). 
\eea
Hence the classical Lagrangian is the Yang-Mills theory in the leading order. 
\\

\noindent
We can also proceed with the extension to all orders by replacing the $\partial$ with the $D$. 
The form of field strength $\hat{{\cal F}}_{\alpha\dot{\mu}}$ remains the same 
\bea
\hat{{\cal F}}_{\alpha\dot{\mu}}\equiv\hat{V}^{-1}{}_{\dot{\mu}}{}^{\dot{\nu}}(\hat{F}_{\alpha\dot{\nu}}+g\hat{F}_{\dot{\nu}\dot{\rho}}\hat{B}_{\alpha}{}^{\dot{\rho}}), 
\eea
but needs the non-Abelian generalization of $\hat{V}_{\dot{\nu}}{}^{\dot{\mu}}$ and $\hat{B}_{\alpha}{}^{\dot{\mu}}$:
\bea
\hat{V}_{\dot{\nu}}{}^{\dot{\mu}}\equiv \delta_{\dot{\nu}}{}^{\dot{\mu}}+gD_{\dot{\nu}}\hat{b}^{\dot{\mu}}; \qquad 
\hat{V}_{\dot{\mu}}{}^{\dot{\nu}}(D^{\alpha}\hat{b}_{\dot{\nu}}-\hat{V}^{\dot{\rho}}{}_{\dot{\nu}}\hat{B}^{\alpha}{}_{\dot{\rho}})
+\epsilon^{\alpha\beta}\hat{F}_{\beta\dot{\mu}}
+g\epsilon^{\alpha\beta}\hat{F}_{\dot{\mu}\dot{\nu}}\hat{B}_{\beta}{}^{\dot{\nu}}=0.
\label{nonA}
\eea 
It is necessary to modify the remaining component of field strength 
\bea
\hat{{\cal F}}_{\alpha\beta}=\hat{F}_{\alpha\beta}
-g(\hat{F}_{\alpha\dot{\mu}}\hat{B}_{\beta}{}^{\dot{\mu}}+\hat{F}_{\dot{\mu}\beta}\hat{B}_{\alpha}{}^{\dot{\mu}})
+\frac{g^2}{2}\hat{F}_{\dot{\mu}\dot{\nu}}(\hat{B}_{\alpha}{}^{\dot{\mu}}\hat{B}_{\beta}{}^{\dot{\nu}}+\hat{B}_{\beta}{}^{\dot{\nu}}\hat{B}_{\alpha}{}^{\dot{\mu}})
\eea
with the anti-symmetrization $\hat{{\cal F}}_{\alpha\beta}=-\hat{{\cal F}}_{\beta\alpha}$. 
All field strengths ($\hat{O}$) are all covariant 
\bea
\hat{\delta}_{\hat{\Lambda}}\hat{O}=i\lbrack\hat{\lambda}, \hat{O}\rbrack+g\hat{\kappa}^{\dot{\mu}}\partial_{\dot{\mu}}\hat{O}. 
\eea
under the gauge transformation: 
\bea
\hat{\delta}_{\hat{\Lambda}}\hat{b}^{\dot{\mu}}&=&\hat{\kappa}^{\dot{\mu}}
+i\lbrack\hat{\lambda}, \hat{b}^{\dot{\mu}}\rbrack
+g\hat{\kappa}^{\dot{\nu}}\partial_{\dot{\nu}}\hat{b}^{\dot{\mu}};
\nn\\
\hat{\delta}_{\hat{\Lambda}}\hat{a}_{\dot{\mu}}&=&\partial_{\dot{\mu}}\hat{\lambda}+
i\lbrack\hat{\lambda}, \hat{a}_{\dot{\mu}}\rbrack
+g(\hat{\kappa}^{\dot{\nu}}\partial_{\dot{\nu}}\hat{a}_{\dot{\mu}}
+\hat{a}_{\dot{\nu}}\partial_{\dot{\mu}}\hat{\kappa}^{\dot{\nu}});
\nn\\
\hat{\delta}_{\hat{\Lambda}}\hat{a}_{\alpha}&=&\partial_{\alpha}\hat{\lambda}+
i\lbrack\hat{\lambda}, \hat{a}_{\alpha}\rbrack
+g(\hat{\kappa}^{\dot{\nu}}\partial_{\dot{\nu}}\hat{a}_{\alpha}
+\hat{a}_{\dot{\nu}}\partial_{\alpha}\hat{\kappa}^{\dot{\nu}}).  
\eea 
We can also read the gauge transformation of the $\hat{B}_{\alpha}{}^{\dot{\mu}}$ from Eq. \eqref{nonA}
\bea
\hat{\delta}_{\hat{\Lambda}}\hat{B}_{\alpha}{}^{\dot{\mu}}=\partial_{\alpha}\hat{\kappa}^{\dot{\mu}}
+i\lbrack\hat{\lambda}, \hat{B}_{\alpha}{}^{\dot{\mu}}\rbrack
+g\big(\hat{\kappa}^{\dot{\nu}}\partial_{\dot{\nu}}\hat{B}_{\alpha}{}^{\dot{\mu}}-(\partial_{\dot{\nu}}\hat{\kappa}^{\dot{\mu}})\hat{B}_{\alpha}{}^{\dot{\nu}}\big). 
\eea
\\

\noindent
After integrating $\hat{b}$, it leaves a determinant depending on $\hat{a}$ in the measure. 
Therefore, it is necessary to concern the determinant for the quantum action. 
The difference between the Abelian and non-Abelian gauge groups also happened in the multiple M5-branes \cite{Chu:2012um}. 
The $\hat{a}_{\alpha}$ in R-R D4-brane theory does not have a direct connection to the NP M5-brane theory \cite{Ho:2011yr}. 
Therefore, we need to dualize a field and then find $\hat{a}_{\alpha}$ \cite{Ho:2011yr}. 
The dualization also leaves the determinant in the measure \cite{Ho:2011yr}. 
The R-R D4-brane theory does not include this quantum contribution \cite{Ho:2011yr}. 
Our construction is the generalization of R-R D-brane theory. 
Hence we only can discuss the classical Lagrangian.

\section{Discussion and Conclusion}
\label{sec:5}
\noindent 
We studied the D-branes with a large R-R ($p-1$)-form field background leading to the discovery of the VPD symmetry and the role of the ($p-1$)-bracket.  
The ($p-1$)-bracket generates the VPD symmetry and this R-R field background in the D-brane \cite{Ho:2011yr}. 
This bracket is closed under the T-duality transformation \cite{Ho:2013paa}, which is a symmetry transformation in String Theory. 
Therefore, the generalization of the bracket should strongly constrain the Lagrangian description.   
We generalized the ($p-1$)-bracket which involves incorporating a non-Abelian one-form gauge field by replacing $\partial$ with $D$. 
The generalized bracket still provides the VPD symmetry and the R-R field background. 
The R-R D-branes were first constructed in Ref. \cite{Ho:2011yr}. 
The generalized bracket reproduces the gauge theory with the VPD symmetry.  
Now, this ($p-1$)-bracket is not closed under the T-duality, so interaction terms between the commutator and bracket are required for multiple D-branes.   
Incorporating a non-Abelianization of the ($p-2$)-form gauge potential introduces complex interaction terms. 
The interaction structure in multiple D-branes is expected to be similar to our computation described. 
The non-Abelianization of the ($p-2$)-form gauge potential can be applied using the generalized bracket due to the R-R field background being restricted to the U(1) sector.  
A Lagrangian description similar to the R-R D-brane theory can be proposed through a modified non-linear equation (replaces $\partial$ with $D$) satisfied by $\hat{B}_{\alpha}{}^{\dot{\mu}}$ and gauge transformation. 
All-order results were obtained, with the gauge transformation matching that of the study of multiple M5-branes described in Ref. \cite{Chu:2012um}. 
The study of multiple M5-branes provides supporting evidence for these findings. 
\\

\noindent
The Ref. \cite{Ma:2020msx} discussed the solution to the SW map for the Abelian $\hat{b}$ case. 
The generalization of the SW map to the case of non-Abelian $\hat{b}$) should not be difficult. 
This implies that the techniques used in the Abelian case can be extended straightforwardly to the non-Abelian case. 
The SW map is a tool used in non-commutative field theory to relate the non-commutative and commutative descriptions of a theory. 
In this case, it helps examine the consistency of the low-energy description from the commutative side \cite{Ma:2020msx}. 
The non-commutative NS-NS D-brane is known to be non-trivial because the quadratic term of field strength contains all-order effects through the Moyal product \cite{Seiberg:1999vs}. 
On the other hand, there is no evidence of similarity to the R-R D-branes. 
Therefore, it would be interesting to apply the SW map to examine the non-commutative R-R D-branes and explore their consistency with the commutative description. 
The SW map provides a useful tool to study these aspects and understand the effects of non-commutativity in the low-energy limit of the theory.  
\\

\noindent 
The most non-trivial direction is the low-energy construction of M2-M5 branes. 
The gauge transformation of $\hat{b}$ is the same as in the study of multiple M5-branes \cite{Chu:2012um}. 
Our construction could be relevant to understanding the higher-order deformation of M5-branes.  
In Ref. \cite{Chu:2012um}, a constraint was introduced to implement self-duality, which resulted in the M5-branes being described as a combination of Yang-Mills theory and a quantum action after performing double dimensional reduction on a circle. 
The appearance of a quantum action in the description of M5-branes is a consequence of the introduced constraint \cite{Chu:2012um}, although the physical origin of this constraint is still unknown. 
Similarly, in the case of R-R D4-brane, a complete description requires considering the quantum contribution from the measure, as discussed in Ref. \cite{Ho:2011yr}. 
Exploring the relationship between the multiple M5-branes and R-R D4-branes, and understanding their quantum behavior, could provide valuable insights into the Lagrangian description of low-energy M5-branes.  

\section*{Acknowledgments}
\noindent
The author would like to thank Chong-Sun Chu, Xing Huang, and Yiwen Pan for their helpful discussion and thank Nan-Peng Ma for his encouragement.
\\

\noindent
The author acknowledges the YST Program of the APCTP; 
Post-Doctoral International Exchange Program (Grant No. YJ20180087); 
China Postdoctoral Science Foundation, Postdoctoral General Funding: Second Class (Grant No. 2019M652926); 
Foreign Young Talents Program (Grant No. QN20200230017); 
Science and Technology Program of Guangzhou (Grant No. 2019050001).

%\appendix

\end{document}